\magnification=\magstep1
\hoffset=-0.6 true cm
\voffset=0.0 true cm
\baselineskip=20pt minus 1pt
\baselineskip=12pt

\vsize=8.9truein
\hsize=6.8truein
\tolerance=10000
\parindent=1truecm
\raggedbottom
\def\pp{\parshape 2 0truecm 5.8truein 2truecm 5.01truein}
\def\ltsima{\ifmmode {\buildrel < \over \sim }
\else
$\buildrel < \over \sim$\fi}
\def\simlt{\lower.5ex\hbox{\ltsima}}
\def\gtsima{\ifmmode {\buildrel > \over \sim }
\else
$\buildrel > \over \sim$\fi}
\def\kms{\ifmmode {\rm \ km \ s^{-1}}
\else
$\rm km \ s^{-1}$\fi}
\def\simgt{\lower.5ex\hbox{\gtsima}}
\def\bline{\hbox to 1 in{\hrulefill}}
\def\etal{{\sl et al.\ }}

\centerline {\bf Two serendipitous low-mass LMC clusters discovered with HST.}

\bigskip
\centerline{Bas\'\i lio X. Santiago$^{1,2}$, Rebecca A. W. Elson$^2$, 
Steinn Sigurdsson$^2$, Gerard F. Gilmore$^2$}
\vskip 0.5 true cm
\centerline {$^1$ Departamento de Astronomia, Universidade Federal do}
\centerline {Rio Grande do Sul, 91501-970, Porto Alegre, RS, Brasil}
\smallskip
\centerline {$^2$ Institute of Astronomy, Cambridge University,}
\centerline {Madingley Road, Cambridge CB3 0HA, United Kingdom}
\bigskip
\centerline{\it Submitted to MNRAS}
\bigskip
\centerline{\bf Key words: Galaxies: Magellanic Clouds, Star clusters.}
\bigskip
\bigskip
{\bf ABSTRACT}
\medskip
We present V and I photometry of two open clusters in the LMC down
to $V \sim 26$.
The clusters were imaged with the {\it Wide Field and Planetary Camera - 2}
on board of the {\it Hubble Space Telescope}, as part of the {\it
Medium Deep Survey} Key-Project. 
Both are low luminosity
($M_V \sim -3.5$), low mass systems ($M \sim 10^3 M_\odot$).
The chance discovery of these two clusters in two parallel WFPC2 fields
suggests a significant incompleteness in the LMC cluster census near the bar.
One of the clusters
is roughly elliptical and compact, with a steep light profile, 
a central surface brightness $\mu_V (0) \sim 20.2~
{\rm mag/arcsec^2}$, half-light radius $r_{hl} \sim 0.9~pc$ (total visual 
major diameter $D \sim 3~pc$) and an estimated mass $M \sim 1500 M_\odot$.
>From the colour-magnitude diagram and isochrone fits we estimate its
age as $\tau \sim 2-5~\times~10^8$ years.
Its mass function has a fitted slope of
$\Gamma = \Delta log \phi (M) / \Delta log M = -1.8 \pm 0.7$ in the
range probed ($0.9 ~\ltsima~ M/M_\odot ~\ltsima~ 4.5$).
The other cluster is more irregular and sparser, having shallower 
density and surface brightness profiles.
We obtain $\Gamma = -1.2 \pm 0.4$, and estimate its mass as 
$M \sim 400 M_\odot$.
A derived upper limit for its age is $\tau ~\ltsima~ 5~\times~10^8$ years.
Both clusters have mass functions with slopes similar to 
that of R136, a massive LMC cluster,
for which HST results indicate $\Gamma \sim -1.2$.
They also seem to be relaxed in their cores and 
well contained in their tidal radii.

\vfill\eject

{\bf 1 INTRODUCTION}
\medskip
The Large Magellanic Cloud (LMC) contains a vast
number of star clusters, with ages varying from 
$10^{7}$ to $10^{10}$ years.
The LMC cluster system is thus suitable for studying the evolution of
physical properties of clusters such as mass, 
radius, light and density profiles,
luminosity functions (LFs) and mass functions (MFs).
These latter may depend not only on age, but also on
metal abundance or environment
and provide relevant information to the physics 
of star formation from fragmenting
gas clouds (McClure \etal 1986, Larson 1991, 1992). 

Derivation of relevant physical parameters in LMC clusters, as well as in
other dense stellar systems, has been limited by lack of spatial resolution.
This situation improved considerably
with the refurbishment of the {\it Hubble Space Telescope} (HST),
allowing crowding problems in dense
stellar systems to be substantially reduced (Elson \etal 1995, de Marchi
\& Paresce 1995a,b, Hunter \etal 1996, Santiago \etal 1996). 
However, the faint end of the cluster luminosity function in the LMC
has not yet been targeted by deep photometric studies.
HST imaging of clusters and
associations published so far 
has concentrated on Galaxy clusters or on rich systems in the Local Group.
Almost nothing is known about
small LMC clusters,
whose detection is often difficult and whose properties
are harder to determine observationally.
The luminosity function of LMC clusters seems to rise
steeply in the low luminosity domain (Elson \& Fall 1985).
In fact, the total number of detected LMC clusters has 
been steadily increasing and is currently
believed to be about $\sim 4500$ (Hodge 1988). 

In this paper, we report observations of two low luminosity
($M_V \sim -3.5$) clusters in the LMC.
They were detected in two parallel {\it Wide Field and Planetary Camera - 2}
(WFPC2) images 
as part of the Medium Deep Survey
HST key-project (MDS). This chance discovery of two clusters suggests
that such objects may be very common in the LMC.
If our two MDS fields are typical, the implied surface density
would be considerably larger than that inferred by Elson \& Fall (1985)
or by the deep photographic study of Hodge (1988), at least in the vicinities
of the LMC bar or the 30 Doradus region.

In \S 2, we present HST photometry in two bands for the two MDS fields;
we show colour-magnitude
diagrams (CMDs) and discuss completeness corrections and
photometric calibration issues.
We then use the data to
extract surface brightness and stellar density profiles for the
clusters, and to estimate their ages (\S 3).
In \S 4, we compute their LFs and MFs.
We briefly discuss their dynamical state as well.
In \S 5, we discuss the results and present our conclusions.

\bigskip 
{\bf 2 THE DATA}
\medskip
{\it 2.1 Data Reduction} 
\medskip
The two LMC fields studied in this work were imaged in parallel mode
with WFPC2, as part of the {\it Medium Deep Survey} (MDS) key
project.
Field 1 is located at $\alpha = 05:35:36.8$ and
$\delta = -69~24~23$ (J2000).
Field 2 is located at $\alpha = 05~36~52.3$ and $\delta = -69~37~59.6$ (J2000).
Both fields are at the eastern end of the LMC bar, southwest
of 30 Doradus, close to NGC2050 and NGC2048, respectively.
Each field contains one of the small clusters 
reported in this work. We hereafter refer to the cluster located in
field 1 (field 2) as C1 (C2).

Field 1 was imaged with two 500s exposures
using the HST F814W (I) filter and two 500s exposures with the HST
F606W (V) filter. For field 2, two exposures in
each filter were taken as well: 4900s and 1100s in F814W and
4900s and 1000s in F606W.

The raw data were processed with the standard pipeline procedure, which
corrects for instrumental effects (Holtzman \etal 1995a).
The two exposures in each field/filter combination were
then coadded and median filtered; the lower instrumental value
was used at each pixel position to eliminate cosmic rays. 
Field 2 F814W frames were offset by about 3'' from each other and
were registered to a common position before coadding. 

\medskip
{\it 2.2 Sample Selection}
\medskip

An object list was obtained separately for each field/filter configuration
from the final coadded image. The IRAF DAOPHOT package (Stetson 1987) was used
for this as well as for aperture and psf fitting photometry.
We adopted a detection threshold of 3 $\sigma$, 
where $\sigma$ is the standard deviation
in the background counts of each chip. 
We worked only with the 3 {\it Wide Field Camera - 2} (WFC2) chips,
since the {\it Planetary Camera} (PC) chip would not 
significantly increase the sample size.

Inspection of the
images showed that most of the objects detected were 
real stars. However,
some spurious detections occurred, especially around bright stars.
In order to clean up the sample, a point spread function (psf) 
template was created from
a few bright unsaturated stars in each WFC2 chip and then fitted to all
remaining objects using ALLSTAR. 
This task gives a $\chi^2$ and a sharpness parameter ($s$). 
The dependence
of these parameters on $V_{606}$ magnitude for chip 4, field 1
is shown in Figure 1. There are two loci in the sharpness diagram:
inspection of the images revealed that the clump of objects with
$s > 0.1$ contains predominantly spurious detections (also attested by 
their large $\chi^2$), whereas the
low sharpness locus was almost entirely made up of stars.
An additional cut in $\chi^2 < 2$ was also applied to eliminate the few
remaining objects whose fit to the psf template was not 
satisfactory. Also, in order to avoid saturation effects,
all objects brighter than $I_{814} = 18$ or $V_{606} = 19$ were
eliminated.
Figure 1 is typical of the other two WFC2 chips in
field 1, both for F606W and F814W.

Field 2 has a longer exposure time, leading to enhanced crowding and
more saturated stars.
The saturation magnitudes measured for field 2 are $I_{814} = 20.5$ and 
$V_{606} = 21.5$. Registration of the F814W images
reduced the quality of the psf fits.
The $s~\times~I_{814}$ and $\chi^2~\times~I_{814}$ diagrams 
did not display as clear boundaries between stellar and non-stellar objects.
Thus, star selection 
for field 2 was restricted to the F606W image,
although object detection was carried
out independently in both filters.
The faint saturation limits in field 2 made it difficult to select
bright stars based on the $\chi^2~\times~ V_{606}$ and $s~\times~ V_{606}$ 
diagrams.
Thus, only stars with $V_{606}$ ~\gtsima~ $21.5$ were selected this way.
Brighter stars were selected
from a smoothed version of the coadded images, obtained by applying  
a 3 pixel ($\sim 0.3''$) Gaussian. This filter is
wide enough to eliminate hot pixels and psf features, but 
narrow enough to allow the detection of bright objects.

The number of stars was typically 6000-7000 per chip in
field 1 and 8000 in field 2.
Of these, about 5000  had {\it both} $I_{814}$ and 
$V_{606}$ magnitudes available. 

\medskip
\centerline {\it 2.3 Photometry}
\medskip
Even though sample selection was mostly based on psf fitting,
the magnitudes and colors used in the analysis came from aperture 
photometry inside a radius $r = 2 ~pix$ ($r = 0.2''$).
$V_{606}$ and $I_{814}$ magnitudes measured this way led to a 
narrower CMD than that based on the psf magnitudes.
The choice of radius is a compromise between the need to bypass centering and
undersampling problems and the need to 
avoid light contamination from neighbouring sources.
An aperture correction of 0.24 mag had to be applied to
both HST filters in order to account for the light outside the
aperture (Holtzman \etal 1995a). 

For field 2, saturation prevented magnitudes and colors of upper main
sequence and the red giant branch (RGB) stars to be measured.
In order to bypass this problem, aperture photometry
for the bright stars in field 2 ($V_{606} < 21.5$) 
was carried out in the F814W and F606W frames
with the shortest exposure time. These have much
less stringent saturation
limits ($I_{814} \sim 18.75$, $V_{606} \sim 19.75$).

Finally, an attempt was made to measure magnitudes for field 2
stars with $I_{814} < 18.75$ or $V_{606} < 19.75$
using only the pixels within the circular
ring between 2 and 3 pixel radius. Again the shortest exposures
were used.
Given the shape of the WFC2's psf, this in principle allows one
to push saturation levels towards brighter magnitudes.
The quality of the magnitudes obtained with this 
procedure was tested with several isolated, bright and 
unsaturated stars. Only approximate magnitudes
could be measured, the uncertainty being
$\sim 0.35$ mag. 
About 50 such bright
stars were added to the sample in each field 2 chip.
\medskip

{\it 2.4 Photometric Calibration} 
\medskip
The data were calibrated to the Johnson-Cousins
system using the ``synthetic'' transformation equations
listed in Table~10 of Holtzman \etal (1995b). 

Reddening corrections were determined by comparing
isochrones and the aperture corrected instrumental
CMDs until the observational and
theoretical main-sequences and RGBs coincided.
We used Yale isochrones for that purpose (Green \etal 1987), and
assumed a distance modulus of $m-M = 18.5$ to the LMC (Panagia \etal 1991).
The Yale isochrones had first to be converted to the HST photometric
system by applying the inverse of the calibration equations listed
by Holtzmann \etal (1995b).
For field 1, a $E(V_{606}-I_{814}) = 0.12$ was obtained.
For field 2 $E(V_{606}-I_{814}) = 0.17$ provided the best fit.
Given the difficulty in fitting all the CMD features, however,
these values may be uncertain by as much as 0.1 mag.

The reddening and aperture corrected 
HST colour-magnitude diagrams (CMDs) for all stars 
in fields 1 and  2
are shown in Figure 2. Also shown are 
Yale isochrones for Z=0.01 stars with ages of 0, 200, 500, 1000 and 
2000$~\times~10^6$ years (Myrs). 
Only non-saturated stars are included.
The CMD for field 1 (panel {\it 2a}) 
shows a main sequence ranging
from $V_{606} \sim 18.5$ down to $V_{606} \sim 26$.
The RGB is also clearly visible in both panels, but especially
for field 1, where
the shorter exposure time and the more accurate sample selection
and photometry allow even the
red giant clump ($V_{606} \sim 18.5-19$) to stand out.
The larger number of objects in Field 2 which lie
outside the main-sequence and RGB reflects the limitation in
star selection, which for field 2 was based on the F606W image only.

The isochrones nicely fit the main sequences
of both panels justifying the adopted reddening and distance modulus.
In particular, the isochrones bracket the width of the upper main sequence, 
which should be made up of fairly young stars and for which the
choice of metallicity is justified (Olszewski \etal 1991).
A detailed analysis and 
discussion of the field stellar populations and star formation history
in these fields was left to Elson \etal (1997).

\medskip
{\it 2.5 Completeness Corrections}
\medskip
Completeness functions were obtained independently for field and
cluster stars, since they are known to depend on crowding.
Completeness functions were measured mostly for WFPC2 chip 4, field 1 and
WFPC2 chip 3, field 2, where C1 and C2 are respectively located.

For the field stars, we ran a total of 
40 realizations of the DAOPHOT.ADDSTAR task
for each HST filter, 5 realizations for each of 8 magnitude bins,
spanning the range $20 \leq I_{814}, V_{606} \leq 28$.
In each realization, 200 stars were added to randomly chosen sections
of the original frame, each section being 400x400 pixels wide and
situated away from the cluster. The section images containing 
the artificial stars were then put through the same detection and star
classification processes as the real data. 
Only artificial stars whose input and observed magnitudes
were within 0.3 mag of each other were considered as detected.
The average fraction,
out of the 5 realizations for each magnitude bin,
of artificial objects detected and classified as stars
was taken as the completeness value
at that magnitude. 

For the cluster regions we made
40 realizations with 10 artificial stars for each
of 12 magnitude bins within the range $20 \leq I_{814}, V_{606} 
\leq 26$. 
Experiments showed that C1 completeness does not depend strongly
on position within the cluster region. This is consistent with the
compact and steeply declining profile for this cluster (\S 3).
For C2, two completeness functions were
assigned, one for its core region and the other for the outskirts.

The $V_{606}$ completeness functions are shown in Figure 3.
Cluster completeness falls more rapidly with
magnitude than field completeness due to more severe crowding.
The differences in completeness between fields 1 and 2
are almost always smaller than the error bars. This applies both
to cluster and field stars completeness functions.

A faint cut-off limit was applied to the data in order to avoid
large shot-noise errors.
Magnitude limits of $I_{814} = 25$ ($I_{814} = 24$) and $V_{606} = 26$
($V_{606} = 25$) were applied to field (cluster) stars. The magnitude
errors (1 $\sigma$) at the cut-off limits 
are $\delta I_{814} \sim 0.2$ for $I_{814} = 25$ and 
$\delta V_{606} \sim 0.3$ for $V_{606} = 26$.
These limits were used in the derivation of structural parameters,
density and surface brightness profiles, LFs and MFs presented 
in \S 3 and \S 4. 

The $I_{814}$ completeness functions behave similarly
to the ones shown in Figure 3. 
Since the final sample used in this work is that made up of objects
with both a $V_{606}$ and a $I_{814}$ magnitude, a joint completeness
function has computed. The additional incompleteness caused by
the requirement of V (I) band detection was quantified
by simply multiplying, at each magnitude level, the I (V) 
completeness function,
by the fraction of sources detected (and
classified as stars) in
I (V) which made into the final sample. 
\bigskip
{\bf 3 THE LOW-MASS LMC CLUSTERS: MORPHOLOGY, STRUCTURE AND AGE}
\medskip
{\it 3.1 Morphology and Structure}
\medskip

$V_{606}$ band images of C1 and C2 are shown in Figures 4 and 5.
C1 is located at $\alpha = 05~35~35.1$ and $\delta = -69~23~48.8$,
whereas C2's centre is at $\alpha = 05~36~51.9$ and
$\delta = -69~38~12.1$ (J2000). These positions are about 20' and 30'
away from the center of the 30 Dor region, respectively.

C1 has visually determined major and minor diameters of $12'' ~\times~ 10''$.
Assuming a distance modulus of $m - M = 18.5$ for the LMC,
this corresponds to $D ~\times~ d~ \sim ~3 ~\times~ 2.5 ~ pc$ (1 pc = 4.1''). 

In Figure 6 we show the stellar number density
(panel {\it a}) and the surface brightness profiles (panel {\it b})
for C1. The upper curve on both panels gives the
profile uncorrected for contamination by field stars.
The error bars include Poisson fluctuations as well as uncertainties
in the completeness functions. The dots show background corrected 
profiles and incorporate 
the additional statistical uncertainty associated with the subtracted
background stars.
Arrows indicate upper limits.
In this case the tip of the arrow is at the
most probable value and its upper end corresponds to
the $1 \sigma$ deviation from this value.
The horizontal lines in both panels indicate the background levels,
determined separately for each WFC2 chip (dotted lines),
and linearly interpolated into the cluster region (solid line).
This latter was used as the best estimate of the field contamination at the
cluster position. The profiles shown include only 
stars in the range $-0.5 \leq M_V \leq 6.5$.

A clear excess of stars is visible out to $R \sim 2~pc$
($\sim 80$ pixels), beyond which the stellar number density
merges with that estimated for the LMC field (panel {\it 6a}).
There are at least as many cluster stars as background ones within this
radius. C1's central regions have a roughly constant
surface brightness, $\mu_V \sim 20.2~ {\rm mag/arcsec^2}$ (panel {\it 6b}).
Beyond $R \sim 0.5~ pc$, however, the surface brightness profile falls 
steeply with radius, flattening out again at $R \sim 2~ pc$.
This outer extension is not visually noticeable (see fig. 4)
and is not present in the number density profile either (panel {\it 6a}).
It could be an artifact caused by
underestimation of background surface brightness levels; 
as panel {\it 6b} itself shows, the background
field $\mu_V$ varies by some 0.3 mag with position within field 1.
However, the uncorrected $\mu_V$ profile 
is still brighter than 
the highest background level estimated from the WFC2 chips.
Since C1 lies in an intersection of several associations and
star forming regions, the excess of surface brightness beyond $R \sim 2~ pc$
may be caused by these larger scale structures in the LMC.
We return to this issue in \S 4.1.

The stellar density and surface brightness profiles for 
C2 are shown in Figure 7. Again only stars with $-0.5 \leq M_V \leq
6.5$ contribute to the profiles.
Because C2 is closer to the chip border than C1, its profiles
do not extend as far from the cluster center as
in the case of C1.
C2 has visual diameters of 19'' x 16'' 
($D ~\times~ d = 4.6~\times~3.9 ~pc$). 
Its visual appearance suggests a sparser and
more irregular cluster showing some substructure.
In spite of the low stellar number
density contrast relative to the background, panel {\it 7a} shows a systematic
excess of stars out to $R \sim 1.5~ pc$, although
this excess is hardly significant beyond $R \sim 1~ pc$.
C2's light profile (panel {\it 7b}), on the other hand, 
is above that of the contaminating field out to $R \sim 1.5~ pc$,
the excess brightness relative to the background being
still significant all the way out to the edge of the chip.
We get $\mu_V (0) \sim 20.3~ {\rm mag/arcsec^2}$ for C2, comparable
to C1.
Its $\mu_V$ profile is shallower than that of C1.

The inferred structural parameters for C1 and C2 are listed in
Table~1, including sizes and central densities.
\medskip
{\it 3.2 Ages}
\medskip
The top panels of Figure 8 show 
the CMD for stars within boxes of 16'' (4 pc) 
on a side centered on C1 (panel {\it a}) and on C2 (panel {\it b}). 
A total of 268 and 310 stars are shown in panels {\it a} and {\it b},
respectively.
The data are in the Johnson-Cousins system and are corrected for 
aperture and extinction effects, as described in \S 2.3 and \S 2.4.
The dashed lines indicate saturation levels. They correspond to a 
fixed $V_{606}$ cut-off (see \S 2.2 and \S 2.3).
The lower panels show CMDs for stars in the neighbourhood of each
cluster, for comparison. The entire area outside the cluster 
in the WFC2 chip where it is located 
was used as comparison field. The field CMDs shown include
a randomly selected fraction of the field stars, so that 
the numbers of cluster and background stars in any region of the CMD
can be directly compared. They clearly differ in the upper 
main sequence: there are 16 main sequence stars 
with $V < 20$ in panel {\it a}, 8 of which have $V < 19$.
In the corresponding comparison field (panel {\it c}), these numbers are
8 and 1. Similarly, only 3 main sequence stars have $V < 20$ in panel 
{\it d}, whereas
in the corresponding cluster CMD there are 14 such stars. Thus,
most of the upper main sequence stars are real cluster members and
we expect field contamination
not to affect age estimates from isochrone fitting.

Yale isochrones corresponding to Z=0.01 stars with ages of 0, 200, 500,
1000 and 2000 Myrs are shown in the upper panels.
This chosen metallicity is typical of LMC clusters (Olszewski \etal 1991). 
Its associated uncertainty ($\delta [Fe/H] \sim 0.15$) has a smaller
effect on the isochrone fits 
than the reddening and saturation effects.
The 500 Myrs isochrone is the one that best fits C1's upper main sequence. 
However, a few saturated stars exist within C1 
and have been left out of Figure 8.
The presence of stars brighter than the 500 Myrs turn-off would indicate
a younger age. In fact, adjustments in the amount of extinction
or in the metallicity assumed for the stars
would allow an age as low as $\tau = 200$ Myrs for C1. 

Assigning an age to C2 is harder, given the larger photometric
errors and stronger saturation effects.
Another problem is that the observed main sequence is a bit redder than the 
isochrones, suggesting a larger reddening within C2's region than elsewhere
in field 2. 
Adjusting the reddening values in order
to match the theoretical and observational MSs and taking into account
the presence of several saturated stars within C2,
we can only set an
{\it upper limit} of $\tau ~\ltsima~ 500$ Myrs to the age of C2.

The derived ages for C1 and C2 are also quoted in Table~1.
In the next two sections we determine the luminosity and mass functions
for both clusters and evaluate their slopes.

\bigskip
{\bf 4 Luminosity and Mass Functions}
\medskip
{\it 4.1 Cluster Luminosity Functions}
\medskip
In Figure 9 we show completeness corrected luminosity functions for C1
and C2.
Field contamination was eliminated by subtracting the field LF from that
within the cluster region. The field LF was taken to be the average
over several control regions equidistant from the cluster.
The cluster regions used for determining their LFs were
circles of 8'' radius centered on each cluster.

C1 seems to have a slightly steeper LF than C2.
LF slopes were obtained from linear fits to the points.
We obtained $\gamma = \Delta log \Phi (M_V) / \Delta M_V = 0.19 \pm 0.03$ 
for C1 in the range $M_V < 6$ and 
$\gamma = 0.12 \pm 0.05$ for C2 in the range $M_V < 4$. 
The best fit lines are shown in the figure. 
Upper limits were not
included in the fits. Given the small number of
LF bins and the low contrast of the clusters, the slope differences
are not significant. In fact, the two LF slopes are similar in the
common range used for the fits ($M_V \leq 3$). C2 LF basically ends beyond
that while C1 LF steepens.
The $\gamma$ values are in agreement with that inferred from the work of
Flower \etal (1980) for NGC1868, a rich LMC cluster with
similar age C1 and C2 but larger mass.
On the other hand, the inferred values for $\gamma$ are smaller
than those typically fitted to younger globular clusters and 
to stellar associations
in the LMC (Vallenari \etal 1993, Will \etal 1995a,b),

In Figure 10, we show C1 LFs for 3 radial bins; the two first are 1 pc wide,
the last is 2 pc wide.
All 3 LFs shown are field subtracted and
were scaled to the entire cluster area,
The number of
stars per unit area decreases with radius,
as expected.
The outermost ring contains only 3 bins in $M_V$ with numbers
significantly above the background. This ring includes the
stars that make
up the excess surface brightness seen beyond $R = 2~ pc$ in panel {\it 6b}.
Despite the uncertainties, the LF looks shallower in panel {\it 10c}
than in {\it 10b}, consistently
with figure 6, where only an excess of light, not stars, is seen.
In the inner rings, on the other hand, the LF
becomes steeper with radius, providing evidence for mass segregation
within C1.
Thus, the bright stars beyond $R \sim 2~ pc$ are likely to be background
stars, belonging either to the general LMC field or
to some stellar association superposed to C1. 
In fact, C1 is situated in a rather messy border region between
different clusters and associations, among them NGC2050, LH96,
DEM261 and NGC157 (see catalogs by Lucke \& Hodge 1970, Davies \etal 1976).
The best fit slopes for the LFs in the two inner rings are $\gamma = 0.09
\pm 0.04$ and $\gamma = 0.25 \pm 0.02$. 
The outer ring lacks enough points for reliable fits to be made.

It was not possible to split C2's LF
into radial sectors, given its smaller contrast with the
background.

\medskip
{\it 4.2 Cluster Mass Functions}
\medskip
Comparing the MFs of clusters with different masses, ages, metallicities
or environments can contribute to the debate about the 
universality of the initial mass function (IMF) and its evolution.

The MF can be derived from the LF with the aid of a mass-luminosity (M-L)
relation, which usually depends on metallicity and is often uncertain.
However, the M-L relation is reasonably well known for 
the typical metallicity of the LMC, $Z \sim 0.008$ ($[Fe/H] \sim -0.3$),
in the range of luminosities covered in this work.
The mass functions for C1 and C2 are shown in Figure 11.
The M-L conversion was based on the Yale isochrone that best
fits the clusters CMDs.
Field subtraction proceeded in the same way as with the LFs. For field stars
we used the same M-L relation as for the clusters. 
The derived mass of the evolved stars will obviously be in error, but these are
subtracted off with the field, having little or 
no effect on the cluster MFs shown.

Linear fits were made to the data points leading to
$\Gamma = \Delta log \phi (M) / \Delta log M = -1.8 \pm 0.7$
and $-1.2 \pm 0.4$ for C1 and C2, respectively.
For C2, however, a single power-law fit
to the MF is inappropriate, since its MF seems to be steeper
for $M ~\gtsima~ 2~ M_\odot$.

The derived slopes are subject to several sources of error. 
Isochrones with different metallicities or based
on stellar models which incorporate convective core overshooting 
would change the M-L relation and therefore the MF slopes.
The effect, however, is known to be small: $\sim 0.1$ in 
$\Gamma$ (Elson \etal 1989, Sagar \etal 1991). 
Unresolved binaries lead to an observed MF which is flatter than
the single star one.
Sagar \& Richtler (1991) investigated this issue and concluded that
the amplitude of the effect is
a function of the binary fraction and the MF slope itself.
The observed slopes derived in this paper may be consistent
with a true $\Gamma \sim -2$ and $-1.5$ for C1 and C2, respectively,
if the fraction of binaries is as large as 0.5. 
Finally, completeness errors may distort the actual shapes of LFs
and MFs, especially if mass segregation is present, as seems to
be the case in C1. Given the small size of the clusters,
it was impossible to split them into many annuli for which
completeness and mass functions could be measured independently.
Given all these uncertainties we cannot rule out a single MF shape
accounting for both of them.

The LF and MF slopes are listed in Table~2, along with
estimates of total luminosities ($M_V$), masses ($M$) 
and mass-to-light ratios ($M/L$). 
These latter were obtained
by adding up the contributions of all non-saturated stars with
$M_V ~\ltsima~ 6.5~(M ~\gtsima~ 0.9~M_\odot)$. 
Thus, the absolute magnitude and mass estimates listed 
should be considered as lower limits. 

The resulting MF slopes are bracketed by most found in the literature.
These were all derived for clusters much larger than the ones
studied here. Elson \etal (1989), for instance, obtained
$-1.8 ~\ltsima~ \Gamma ~\ltsima~ -0.8$ for six rich young
($\tau \sim 30-50~$ Myrs) LMC clusters in the
mass range $1.5 ~\ltsima~ M/M_\odot ~\ltsima~ 6$.
Such flat IMFs were in large contrast with the results
of Mateo (1988), who found
slopes ($-2.5 ~\ltsima~ \Gamma ~\ltsima~ -4.6$) for
six LMC clusters spanning a wider range of ages and metallicities.
Sagar \& Richtler (1991) derived intermediary slopes, 
$\Gamma \sim -2.1$, for five LMC clusters within
the mass range $2 < M/M_\odot < 14$ and spanning an age range
of $10-100~~$ Myrs.

More recently and using HST data, 
Hunter \etal (1995) obtained $\Gamma = -1.22$ for R136, 
a $\sim 2 ~\times~ 10^4~M_\odot$, very young and compact system
in the
center of 30 Doradus, in the range $2.8 < M/M_\odot < 15$. 
Our results suggest flat MF slopes for low mass LMC clusters as well.

A more meaningful comparison would be 
with open clusters in the Galaxy.
Ground-based studies include that of Sagar \etal (1986)
who derived MF slopes for 11 open clusters.
One of them (N1778C) has an age comparable to those of C1 and C2 in
\S 3.2. Its slope is $\Gamma = -2.04 \pm
0.44$ in a mass range comparable to ours. This is consistent with
C1 and marginally consistent with C2.
Other more recent ground-based
works have led to shallower slopes, $\Gamma ~\gtsima~ -1$, but usually in the
mass range $M ~\ltsima~ 0.5~M_\odot$ (Comer\'on
\etal 1996, Hambly \etal 1995 and references therein). 

\medskip
{\it 4.3 Internal dynamics}
\medskip
We now use the integrated properties and structural parameters listed
in Tables~1 and 2 to assess the importance of dynamical
effects such as two body relaxation and tidal forces.
Knowledge of the internal dynamical state of C1 and C2
may help determining the extent to which 
such low mass clusters could contribute to the
field population.

Using the LMC tidal field derived by Elson, Fall \& Freeman (1987),
we estimate C1 and C2 to have similar tidal radii, in the range
$6 ~\ltsima~ r_{t}~(pc) ~\ltsima~ 15$, implying $r_t > r_{vis}$ (see Table~1).
Therefore, the clusters are so far not tidally truncated.
Both, however, may 
have experienced stronger tidal fields or had close 
encounters with other more massive clusters
or associations during their $~\gtsima~ 10^8$ years of existence. 

Is two-body relaxation relevant to the internal dynamics of C1 and C2?
The relaxation time scale for stars with a mass $M$ at a
distance $r$ from the centre is given by

$$t_{rel} = { {2~\times~10^8} \over {ln \Lambda} }~ \Bigl( {r \over {2~pc}} 
\Bigr)^2~ \Bigl( {v \over {0.2~\kms}} \Bigr)~ 
\Bigl( { {M_\odot} \over M} \Bigr), \eqno (1) $$

where $v$ is the typical velocity of such a star
within the cluster.
Assuming hydrostatic equilibrium in the central regions, the
central dispersion velocity in the cluster is given by

$$\sigma^2 (0) \sim G~\rho_0~r_c^2, \eqno (2)$$

where $r_c$ is the core radius and $\rho_0$ is the central mass
density. Using the relation between projected and spatial
densities given by Djorgovski (1993), we estimate 
$\rho_0 \sim 16~ {\rm M_\odot/pc^3}$
for C1 and $\rho_0 \sim 6~ {\rm M_\odot/pc^3}$ for C2. 
These values for the central densities
imply $\sigma (0) \sim 0.4~\kms$ for C1 and
$\sigma (0) \sim 0.2~\kms$ for C2. Inserting these velocities
into equation (1) we infer that $t_{rel} ~\ltsima~ 100$ Myrs for a
1 $M_\odot$ star in the central regions of both clusters. Therefore,
both C1 and C2 should be relaxed in their central parts.
The evidence for mass segregation in C1 is consistent with that.
Dispersion velocities 
just about twice the values inferred would disrupt the clusters
in $\ltsima~50~Myrs$. Given their estimated ages, 
C1 and C2 are probably bound.

\bigskip

{\bf 5 DISCUSSION AND CONCLUSIONS} 
\medskip
We presented V and I photometry of two WFPC2 fields located
near the eastern end of the LMC bar.
Stars as faint as $M_V \sim 7~(M \sim 0.8 M_\odot)$
were detected in each of them.
Each field contained one 
small open cluster, implying a large number density of such 
systems in the LMC.

C1 is roughly symmetrical in shape, has a steep density profile
and a mass function slope of $\Gamma = -1.8 \pm 0.7$. 
>From isochrone fits, we infer an age of $\tau \sim 200-500~$ Myrs
for it. Its estimated mass is $\sim 1500~M_\odot$
and its absolute magnitude, $M_V \sim -3.5$. 
The derived values for the luminosity, mass, MF slope and
age are mutually consistent in light of the recent stellar
population synthesis models of Girardi \etal (1995). These authors
use the photometric models of single stellar populations
calculated by Bertelli \etal (1994) in order to revise the
relation between integrated photometric properties, age and metallicity
of LMC clusters.
>From their Figure 13, we infer that a $M \sim 2~\times~10^3 M_\odot$ 
cluster with
a MF slope a bit shallower than a Salpeter one
($\Gamma = -2.35$; Salpeter 1955) would have $M_V \sim -3.5$ 
at an age $\tau \sim 10^8$ years.

Marginal evidence of mass segregation within $r ~\ltsima~ 2~pc$
from C1's centre was found.
This is consistent with the short relaxation time scale expected for
$M ~\gtsima~ 1~ M_\odot$ stars in its central regions.
Tidal effects from the LMC should not be relevant;
we derive $r_{hl} ~\ltsima~ 1~ pc$ from C1's surface brightness profile, 
which places its member stars well inside the estimated tidal radius 
($8~\ltsima~ r_t~(pc) ~\ltsima~ 15$).
Given the central density and size estimates for C1, its stars should
have a maximum central velocity dispersion of $\sim 0.4 \kms$.
It would be interesting to confirm that
with observations.

C2 seems less massive but just as luminous as C1 ($M \sim 400 M_\odot$,
$M_V \sim -3.5$). This implies flatter luminosity and mass functions.
We obtain $\Gamma = -1.2 \pm 0.4$ for C2. For $M_V ~\ltsima~ 3.5$, C2's LF
is similar to C1's but it drops off for fainter magnitudes.
C2 is more irregular than C1 and has shallower density and surface brightness
profiles. We could only derive 
an upper limit of $\tau ~\ltsima~ 500~$ Myrs to the age of C2 based on its
CMD due to saturation and reddening uncertainties.
Based on the results of 
Girardi \etal (1995) and our mass, $M_V$ and $\Gamma$ estimates,
we would obtain more stringent limits: $\tau ~\ltsima~ 100 Myrs$.
C2's size was harder to quantify, given its flat profile. However, 
just as C1, it is also likely to be contained within its tidal radius
and to have undergone significant core relaxation: $t_{rel} \sim 100$ Myrs
for $M ~\gtsima~ 1 M_\odot$ stars within the central 2 pc.

The chance discovery of two small
clusters in two MDS fields within the LMC 
suggests that these could be more common in the LMC bar or 30 Doradus region
than previously anticipated.
Such systems may have significantly contributed 
to the field star population if
they were even more common in the past and got disrupted by larger clusters
or by the LMC tidal field. Even though that does not seem to be the
case for C1 and C2, other similar systems may have been subjected to stronger
tidal fields from both the LMC or its bar or from more massive clusters or
associations. Alternatively many such clusters might have been unbound 
since their birth, their member stars streaming away after a few hundred Myrs.
It would interesting to confirm or not the
existence of a large population of low mass and luminosity clusters in
the LMC with other observations using the high resolution of HST.

\medskip\nobreak
\bigskip
{\bf ACKNOWLEDGMENTS} 
\medskip\nobreak
We thank Sally Oey and Eduardo Bica for their help
and useful discussions. BXS is grateful to the hospitality
of the Institute of Astronomy, where part of this
work was carried out.
\vfill\eject

{\bf REFERENCES}
\def\pp{\parshape 2 0truecm 13.4truecm 2.0truecm 11.4truecm}
\def\apjref #1;#2;#3;#4 {\par\pp#1, #2, #3, #4 \par}

\hyphenation{MNRAS}
\medskip

\pp Bertelli, G. Bressan, A., Chiosi, C., \& Fagotto, F., 1994, 
{\it A\&A Suppl.}, 106, 275

\pp Comer\'on, F., Rieke, G.H., \& Rieke, M.J., 1996, {\it ApJ}, 473, 294.

\pp Davies, R.D., Elliot, K.H., \& Meaburn, J., 1976, {\it MmRAS}, 81,
89

\pp de Marchi, G., \& Paresce, F., 1995a, {\it A\&A}, 304, 202

\pp de Marchi, G., \& Paresce, F., 1995b, {\it A\&A}, 304, 211

\pp Djorgovski, S., 1993, in {\it The Structure and Dynamics of
Globular Clusters}, PASP Conf. Ser. 50, p. 373

\pp Elson, R.A.W., \& Fall, S.M., 1985, {\it PASP}, 97, 594

\pp Elson, R.A.W., Fall, S.M., \& Freeman, K.C., 1987, {\it ApJ}, 323, 54

\pp Elson, R.A.W., Fall, S.M., \& Freeman, K.C., 1989, {\it ApJ}, 336, 734

\pp Elson, R.A.W., Gilmore, G., Santiago, B.X., \& Casertano, S., 1995, AJ,
110, 682

\pp Elson, R.A.W., Gilmore, G.F., \& Santiago, B.X., 1997, {\it MNRAS}, 
in press

\pp Flower, P.J., Geisler, D., Hodge, P., \& Olszewski, E. W., 1980, {\it ApJ},
235, 769

\pp Girardi, L., Chiosi, C., Bertelli, G., \& Bressan, A., 1995, {\it A\&A}, 
298, 87

\pp Green, E.M, Demarque, P., \& King, C. R., 1987, Yale University 
Transactions, Yale Univ. Observatory

\pp Hambly, N.C., Steele, I.A., Hawkins, M.R.S., \& Jameson, R.F., 1995, 
{\it MNRAS}, 273, 505

\pp Holtzmann, J.A., \etal, 1995a, {\it PASP}, 107, 156

\pp Holtzmann, J.A., \etal, 1995b, {\it PASP}, 107, 1065

\pp Hunter, D.A., Shaya, E.J., Holtzman, J.A., Light, R.M., 1995, {\it ApJ},
448, 179

\pp Hunter, D.A., O'Neil, E.J., Lynds, R., Shaya, E.J., Groth, E.J.,
\& Holtzman, J.A., 1996, {\it ApJ Letters}, 459, L27

\pp Larson, R.B., 1991, in {\it Fragmentation of Molecular Clouds and
Star Formation}, IAU Symp 147, eds: E. Falgarone, F. Boulanger, G. Duvert.
Kluwer, Dordrecht, p. 261

\pp Larson, R. B., 1992, MNRAS, 256, 641.

\pp Lucke, P.B., \& Hodge, P.W., 1970, {\it AJ}, 75, 171

\pp Mateo, M., 1988, {\it Apj}, 331, 261

\pp McClure, R.D., VandenBerg, D.A., Smith, G.H., Fahlman, G.G, Richer,
H.B., Hesser, J.E., Harris, W.E., Stetson, P.B., Bell, R.A., 1986, 
{\it ApJ}, 307, L49

\pp Olszewski, E., Schommer, R., Suntzeff, N. \& Harris, H. 1991,
{\it AJ}, 101, 515

\pp Panagia, N., Gilmozzi, R., Macchetto, F., Adorf, H.-M., 
\& Kirshner, R., 1991, {\it ApJ Letters}, 380, L23

\pp Sagar, R., \& Richtler, T., 1991, {\it A\&A}, 250, 324.

\pp Salpeter, E.E., 1955, {\it ApJ}, 121, 161

\pp Santiago, B.X., Elson, R.A.W., \& Gilmore, G.F., 1996, MNRAS, 281, 1363

\pp Stetson, P.B., 1987, PASP, 99, 191.

\pp Vallenari, A., Bomans, D.J., de Boer, K.S., 1993, {\it A\&A}, 268, 137.

\pp Will, J.-M., V\'azquez, R.A., Feinstein, A., \& Seggewiss, W., 1995a,
{\it A\&A}, 301, 396.

\pp Will, J.-M., Bomans, D.J., de Boer, K.S., 1995, {\it A\&A}, 295, 54.

\vfill\eject

\centerline {Table 1. Clusters structural parameters.}
\vskip 1.5 true cm
\tabskip=1em plus2em minus .5em
\halign to\hsize
{#\hfil&&\hfil#&\hfil#\cr
\noalign {\medskip \hrule \medskip}
Parameter & ~~~~~~~~C1~~~~~~~~ & ~~~~~~~~C2~~~~~~~~ \cr
\noalign {\medskip \hrule \medskip}
$\sigma_0~{\rm (stars/pc^2)}$ & $30 \pm 8$ & $8 \pm 6$ \cr
$R_{\sigma}~{\rm (pc)}$ & $\sim 2$ & $\sim 1.5$ \cr
$\mu_V (0)~{\rm (mag/arcsec^2)}$ & $20.2 \pm 0.4$ & $20.3 \pm 0.5$ \cr
$R_{\mu}~{\rm (pc)}$ & $\sim 2$ & $\sim 2$ \cr
$R_{hl}~{\rm (pc)}$ & $ \sim 0.9$ & $\sim 0.7$ \cr
$D_{vis}~\times~ d_{vis}~{\rm (pc)}$ & $2.9 ~\times~ 2.4$ & $4.6 ~\times~ 3.9$ \cr
$\tau (Myrs)$ & 200-500 & $~\ltsima~ 500$ \cr
}
\medskip
\hrule
\medskip

\vskip 1.5 true cm

\vfill\eject

\centerline {Table 2. LF and MF slopes.}
\vskip 1.5 true cm
\tabskip=1em plus2em minus .5em
\halign to\hsize
{#\hfil&&\hfil#&\hfil#\cr
\noalign {\medskip \hrule \medskip}
Parameter & ~~~~~~~~C1~~~~~~~~ & ~~~~~~~~C2~~~~~~~~ \cr
\noalign {\medskip \hrule \medskip}
$\gamma~=~\Delta log \Phi(M_V) / \Delta log M_V$ & $0.19 \pm 0.03$ & $0.12 \pm 0.05$ \cr
$\Gamma~=~\Delta log \phi(M) / \Delta log M$ & $-1.8 \pm 0.7$ & $-1.2 \pm 0.4$ \cr
$M_V^1$ & $\sim -3.5$ & $\sim -3.5$ \cr
$Mass^1$~($M_\odot$) & $\sim 1500$ & $\sim 400$ \cr
$M/L~(M_\odot/L_\odot)$ & $\sim 0.7$ & $\sim 0.2$ \cr
}
\medskip
\hrule
\medskip
$^1$ Due to stars in the range $-0.5 ~\ltsima~ M_V ~\ltsima~ 6.5~
(0.9 ~\ltsima~ M/M_\odot ~\ltsima~ 4.5)$ only.

\vskip 1.5 true cm

\vfill\eject

\centerline {FIGURE CAPTIONS}
\medskip
\item {1-} The sharpness ($s$) and 
$\chi^2$ parameters, as obtained by fitting each detected object in 
chip 4, field 1
to the WFC2 psf, are shown as a function of $V_{606}$ magnitudes.
{\it a)} $s$. {\it b)} $\chi^2$.

\item {2-} {\it a)} CMD for all 3 WFC2 chips in 
field 1. The magnitudes and colours are in the HST WFPC2 system.
The data are corrected for aperture and reddening effects.
Superposed are Yale isochrones corresponding to Z=0.01 stars with
ages 0, 200, 500, 1000 and 2000 Myrs. 
{\it b)} The same as in $a$ but now for the 3 WFC2 chips in field 2.

\item {3-} {\it a)} $V_{606}$ completeness functions
for field stars in field 1, chip 4 (squares and dashed line) and in
field 2, chip 3 (crosses and dotted line). Error bars
are standard deviations from different realizations of
ADDSTAR, as explained in the text. {\it b)} Same as in
$a$ but now cluster completeness functions are shown: squares and
dashed line: C1; crosses and dotted lines: inner and outer C2. Error bars
are shown only for one of the cluster completeness functions to avoid 
confusion.

\item {4-} WFPC-2 image of LMC field 1 cluster,
 C1. The whole field shown is about 50'' on a side.

\item {5-} WFPC-2 image of LMC field 2 cluster, 
C2. The whole field shown is about 40'' on a side.

\item {6-} {\it a)} Stellar density profile within the
region occupied by C1. The upper curve gives the surface density
of stars as a function of distance from the cluster centre, without
any correction for field contamination. The points indicate
the background corrected profile. Arrows correspond to upper limits.
The adopted background density level is
given by the solid horizontal line. The dotted horizontal lines show
the background densities estimated separately for each WFC2 chip.
{\it b)} V band surface brightness profiles and background levels corresponding
to the number density ones shown in panel {\it a}.

\item {7-} {\it a)} 
The same as in panel {\it 6a} but now
for C2.
{\it b)} The same as in panel {\it 6b} but now
for C2.

\item {8-} {\it a)} The colour magnitude diagram of
stars within C1 region. 
Yale isochrones of 0, 200, 500, 1000 and 2000 Myrs
have been superposed to the data. 
The dashed line indicates the
saturation level.
{\it b)} The same as in panel {\it a}
but now for C2. {\it c)} CMD for field comparison region close to C1.
{\it d)} CMD for field comparison region close to C2.

\item {9-} The luminosity functions of C1 and C2 are shown as solid
and open squares, respectively.
The error bars incorporate Poisson fluctuations as well as 
uncertainties in the completeness corrections and in the background
subtraction. Upper limits are shown with arrows. The lines show
power-law fits to the data.

\item {10-}  The luminosity function for C1
in different radial annuli, as indicated.
The numbers in all panels are scaled to the total cluster
area. Error bars again incorporate
statistical uncertainties as well as completeness and background
errors. Arrows correspond to upper limits. The solid line shown
in the two upper panels is the best fit to a power-law.

\item {11-}  Mass functions of C1 (solid symbols)
and C2 (open symbols).
The error bars incorporate Poisson fluctuations as well as 
uncertainties in the completeness corrections and in the background
subtraction. The lines show the best fit power-laws in each case
(C1: solid line; C2: dashed line).

\vfill\eject
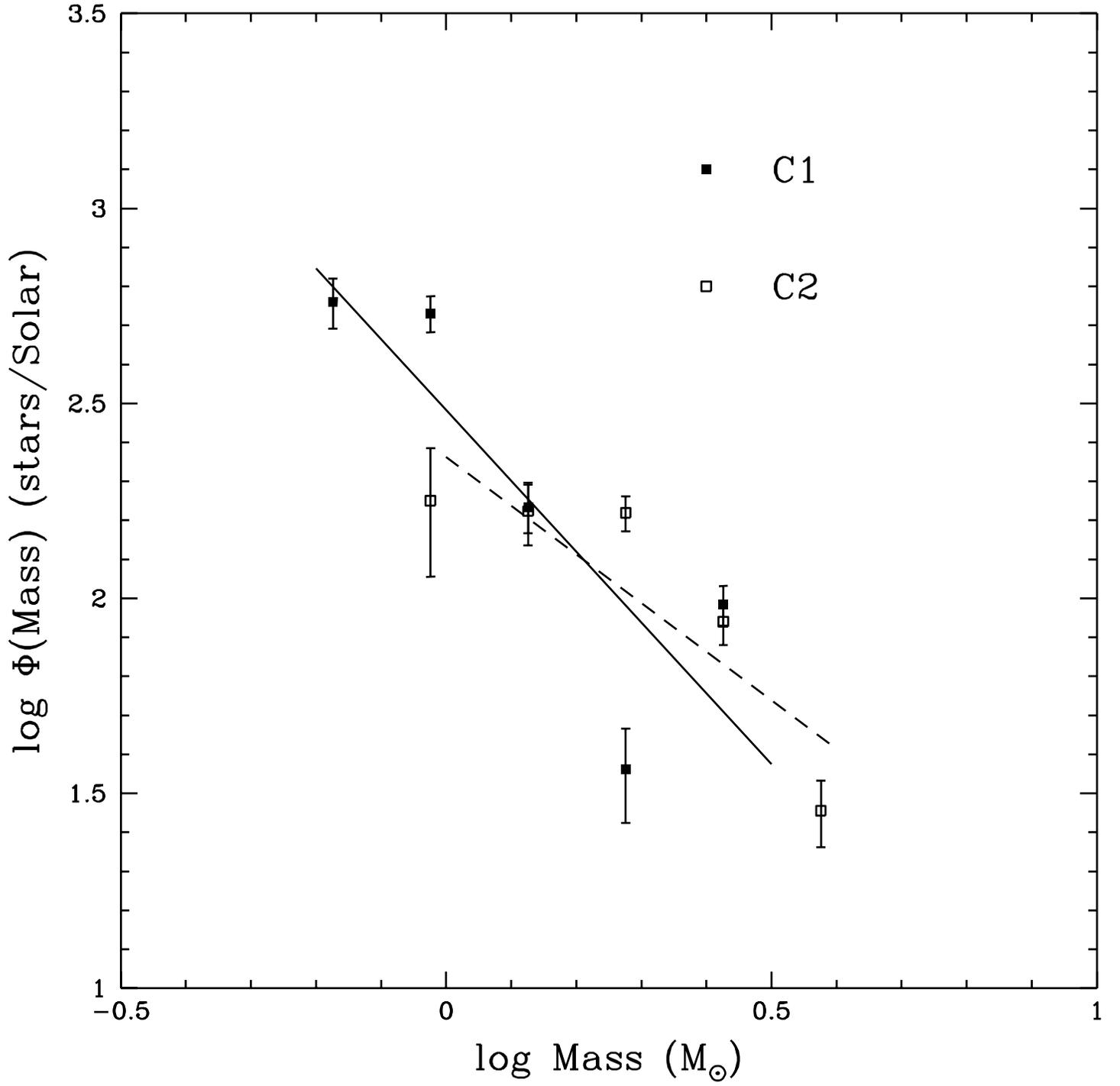
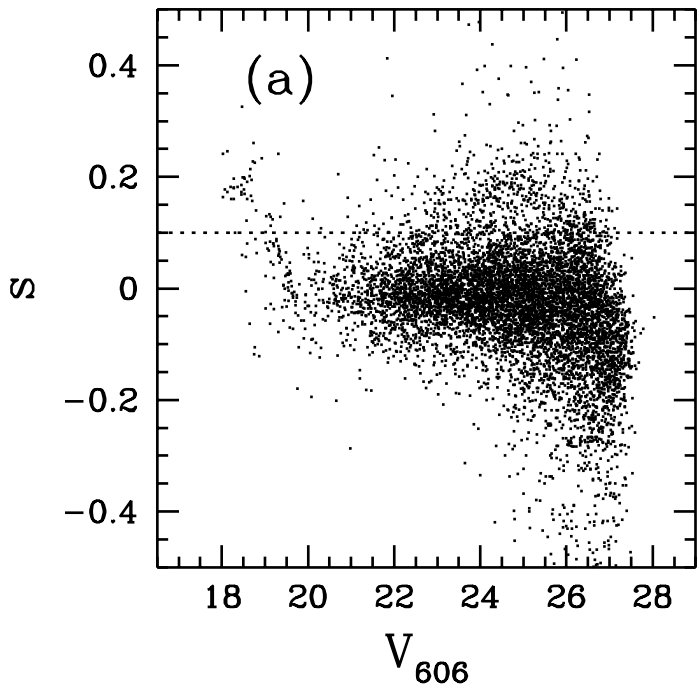
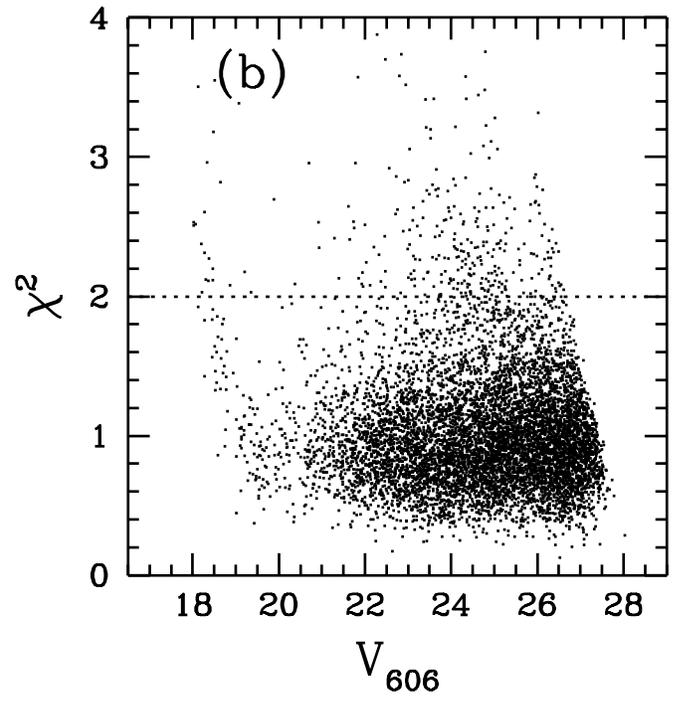
\end